\documentclass[aps,prf]{revtex4}{

\usepackage{amsmath,amsfonts,bm, color,amssymb,float}
\usepackage{graphicx,epsfig,overpic,hyperref,ulem,caption,subcaption}

\usepackage{mathrsfs}

\newcommand{\eps}{{\varepsilon}}

\newcommand{\ehd}{{\mathrm{ehd}}}

\newcommand{\e}{{\hat{\bf e}}}

\newcommand{\bs}{\boldsymbol}

\newcommand{\bE}{{\bs{E}}}

\newcommand{\bI}{{\bf I}}

\newcommand{\bT}{{\bf T}}

\newcommand{\bv}{{\bs{u}}}
\newcommand{\bu}{{\bs{u}}}

\newcommand{\bff}{{\bs{f}}}

\newcommand{\bn}{{\bf n}}
\newcommand{\bsigma}{{\bs{\sigma}}}

\newcommand{\scrL}{\mathscr{L}}
\newcommand{\scrH}{\mathscr{H}}
\newcommand{\scrF}{\mathscr{F}}

\newcommand{\calL}{\mathcal{L}}
\newcommand{\calR}{\mathcal{R}}
\newcommand{\calW}{\mathcal{W}}
\newcommand{\calP}{\mathcal{P}}
\newcommand{\calS}{\mathcal{S}}

\newcommand{\sfP}{\mathsf{P}}
\newcommand{\sfH}{\mathsf{H}}

\newcommand{\bnabla}{\bn{\nabla}}

\newcommand{\refeq}[1]{Eq. (\ref{#1})}
\newcommand{\reffig}[1]{Fig. (\ref{#1})}

\newcommand{\1}{{(1)}}
\newcommand{\0}{{(0)}}

\begin{document}

\title{Electrohydrodynamic flow about a colloidal particle suspended in a non-polar fluid}

\author{Zhanwen Wang$^{1}$,
 Michael J. Miksis$^{2}$
 and Petia M. Vlahovska$^{2}$ {\email{petia.vlahovska@northwestern.edu}}}

\affiliation{{1} Theoretical and Applied Mechanics Program, Northwestern University, Evanston, IL 60208 USA\\
{2} Engineering Sciences and Applied Mathematics, Northwestern University, Evanston, IL 60208 USA}

\begin{abstract}
Nonlinear electrokinetic phenomena, where electrically driven fluid flows depend nonlinearly on the applied voltage, are commonly encountered in aqueous suspensions of colloidal particles. A prime example is the induced-charge electro-osmosis, driven by an electric field acting on diffuse charge induced near a polarizable surface. Nonlinear electrohydrodynamic flows also occur in non-polar fluids, driven by the electric field acting on space charge induced by conductivity gradients. Here, we analyze the flows about a charge-neutral spherical solid particle in an applied uniform electric field that arise from conductivity dependence on local field intensity. The flow pattern varies with particle conductivity: while the flow about a conducting particle has a quadrupolar pattern similar to induced-charge electro-osmosis albeit with opposite direction, the flow about an insulating particle has a more complex structure. We find that this flow induces a force on a particle near an electrode that varies non-trivially with particle conductivity: while it is repulsive for perfectly insulating particle and particles more conductive than the suspending medium, there exists a range of particle conductivities where the force is attractive.
The force decays as inverse square of the distance to the electrode and thus can dominate the dielectrophoretic attraction due to the image dipole, which falls off with the fourth power with the distance. This electrohydrodynamic lift opens new possibilities for colloidal manipulation and driven assembly by electric fields.
\end{abstract}

\date{ \today}

\maketitle

\section{Introduction}

The interaction of colloids and electric fields is widely used for directed assembly and particle manipulation \citep{PRIEVE:2010,Velev:2006,Blaaderen:2013,Edwards:2014,Velev_review:2015, Harraq:2022}. 
In recent years, motile colloids energized by an applied electric field have become a popular model for self-propelled ``active" particles \citep{Yan:2016,Han:2018,Driscoll:2019,Diwakar:2022,BOYMELGREEN:2022}. One propulsion mechanism exploits the induced charge electrophoresis of colloids suspended in aqueous electrolyte solutions  \citep{squires2004induced, Squires-Bazant:2006, Velev:2008, Ma-Wu:2015, Nishiguchi:2015}. Another propulsion strategy is particle rolling on an electrode surface due to the Quincke rotation (a symmetry-breaking instability which gives rise to a torque on the particle in an applied uniform electric field)  \citep{Bartolo:2013, Bartolo:2015,Snezhko:2016,Karani:2019,Gerardo:2019,Zhang:2021a}. The threshold for the Quincke rotation is very sensitive to the solvent conductivity and experimentally accessible only in non-polar solvents, still at electric fields with magnitude in the order of MV/m.  At such strong electric fields, electric conduction may be no longer be in the Ohmic regime due to fluid conductivity becoming dependent on electric field intensity \citep{Onsager:1934,Castellanos:book}. Field-enhanced conductivity arises from the electric field effect on the dissociation-recombination equilibrium between ion pairs and free ions (termed Onsager effect). In a non-polar fluid, the electrolyte added to control conduction exists mostly in the form of neutral ion pairs \citep{PRIEVE:2017}. The application of a strong electric field 
increases the rate of the ion pair dissociation thereby increasing the number of charge carriers and, accordingly, the electrical conductivity \citep{Castellanos:book}.
This effect is suggested to underlie the flow observed about colloids suspended in oil \citep{ryu2010new}, whose pattern resembles the induced charge osmotic flow about an ideally-polarizable particle in aqueous solutions, and the oscillatory motion of Quincke rollers \citep{zhang2021quincke}. Recent experiments have also reported  that a charge-free, dielectric particle  lifts off from the electrode \citep{Gerardo:2019} despite 
the attraction by the image dipole, which may involve electrohydrodynamic flow. Motivated by the potential impact of electrohydrodynamic flows on Quincke colloid ``activity" and collective dynamics, here we examine the possibility of a flow driven by  conductivity gradients set by nonuniformities in the applied electric field.  While the electric field driven flows about colloids near electrodes in aqueous electrolyte solutions have been subject to a great interest \citep{ristenpart2004assembly,ristenpart2007electrohydrodynamic,Hashemi:2018,Bazant:2009,PRIEVE:2010,Khair:2022,Katzmeier:2022,Mateo:2022}, colloidal electrohydrodynamics in non-polar fluids is far less explored. 

In this paper,  we predict that  an electrohydrodynamic flow driven by the Onsager effect arises about a spherical particle in an applied uniform electric field. We develop an asymptotic solution in the case of fluid conductivity linearly varying with the electric field intensity. We  analyze the flow effect on the particle interaction with the electrode. The force on the particle due to the electrohydrodynamic flow is calculated using the Lorentz reciprocal theorem  and found to be repulsive for insulating particles.

\section{Problem formulation}\label{sec:Formulation}
Let us consider a non-polar liquid, e.g., hydrocarbon oil, containing an electrolyte, e.g., Tetrabutylammonium Bromide. In such solutions, the electrolyte exists mostly in the form of neutral ion pairs resulting in very low electric conductivity (in contrast, in aqueous solutions ``strong'' electrolytes are completely ionized). The leaky dielectric model was developed to describe the flows is such weakly conducting fluids \citep{melcher1969electrohydrodynamics,Saville:1997,Vlahovska:2019} adopting Ohm's law for the electric current, whose conservation at steady state results in
\begin{equation}
\label{LDMc}
\nabla\cdot\left(\sigma_m \bE\right)=0\,.
\end{equation}
If the fluid conductivity is constant \refeq{LDMc} implies that the bulk fluid is electroneutral. Charge accumulates only at interfaces separating media with different electric properties.
A field-dependent conductivity $\sigma (E)$ due to field-enhanced electrolyte ionization \citep{Onsager:1934} gives rise to space charge in a spatially inhomogeneous electric field as seen from the conservation of current  \refeq{LDMc} and  the Gauss' law, $\eps_m \bnabla \cdot \bE = \rho_f$
\begin{equation}
	\rho_f = -\frac{\eps_m}{\sigma_m}\bE \cdot \bnabla\sigma_m \,.
 \label{rhof}
 \end{equation}
 The induced charge  in the bulk would then drive flow, which in the creeping flow limit is described by the Stokes equation
\begin{equation}
-\bnabla p + \mu \nabla^2 \bu =-\rho_f \bE,\quad ~ \bnabla \cdot \bu = 0,
\end{equation}
where $\bv$ and $p$ are the fluid velocity and pressure, and $\mu$ is the fluid viscosity.  

In this study, we ask the question ``If a particle is introduced in a uniform electric field, would the resulting field inhomogeneities give rise to  conductivity gradients and space-charge-driven flow? What is the correction due to the Onsager effect to the behavior predicted by the leaky dielectric model, which is no flow about a solid particle?

\subsection{The Onsager effect}
If charge injection is negligible  \citep{DENAT1982197, Sainis:2008,park2009effect}, conduction is due to ions produced from the dissociation of the ion pairs \citep{Castellanos:book,PRIEVE:2017}. In strong fields, the dissociation rate increases with field intensity \citep{Onsager:1934}. The activation energy for the ionization includes the Born self-energy of the two charged ions, the Coulomb
energy of interaction between them, and the energy of separating the charged pair in the external electric field. The latter contribution leads to the  fluid conductivity increasing as \citep{Onsager:1934,Castellanos:book}
\begin{equation}
\label{onsager}
\sigma_m= \sigma_0 F(b)^{1/2}
\end{equation}   
where $\sigma_0$ is the zero-field conductivity, and $F(b)$ is the Onsager function 
\[
 F(b)=\frac{I_1(2b)}{b}\,,\quad b=\left(\frac{e^3 E}{4 \pi \eps_m (k_BT)^2}\right)^{1/2}\,.
\]
Here $I_1$ is the modified Bessel function, $e$ is the electron charge, $k_BT$ is the thermal energy, $\eps_m$ is the fluid permittivity, and $E$ is the field intensity.
\refeq{onsager} shows that conductivity is modified by the presence of an external electric field only if the field intensity is high, typically exceeding MV/m. If $b\ll 1$, $\sigma_m=\sigma_0$ is field-independent and the electrohydrodynamic flow is described by the leaky dielectric model \citep{melcher1969electrohydrodynamics}.

\subsection{Governing equations}

\begin{figure}
    \centering
    \includegraphics[width=0.8\linewidth]{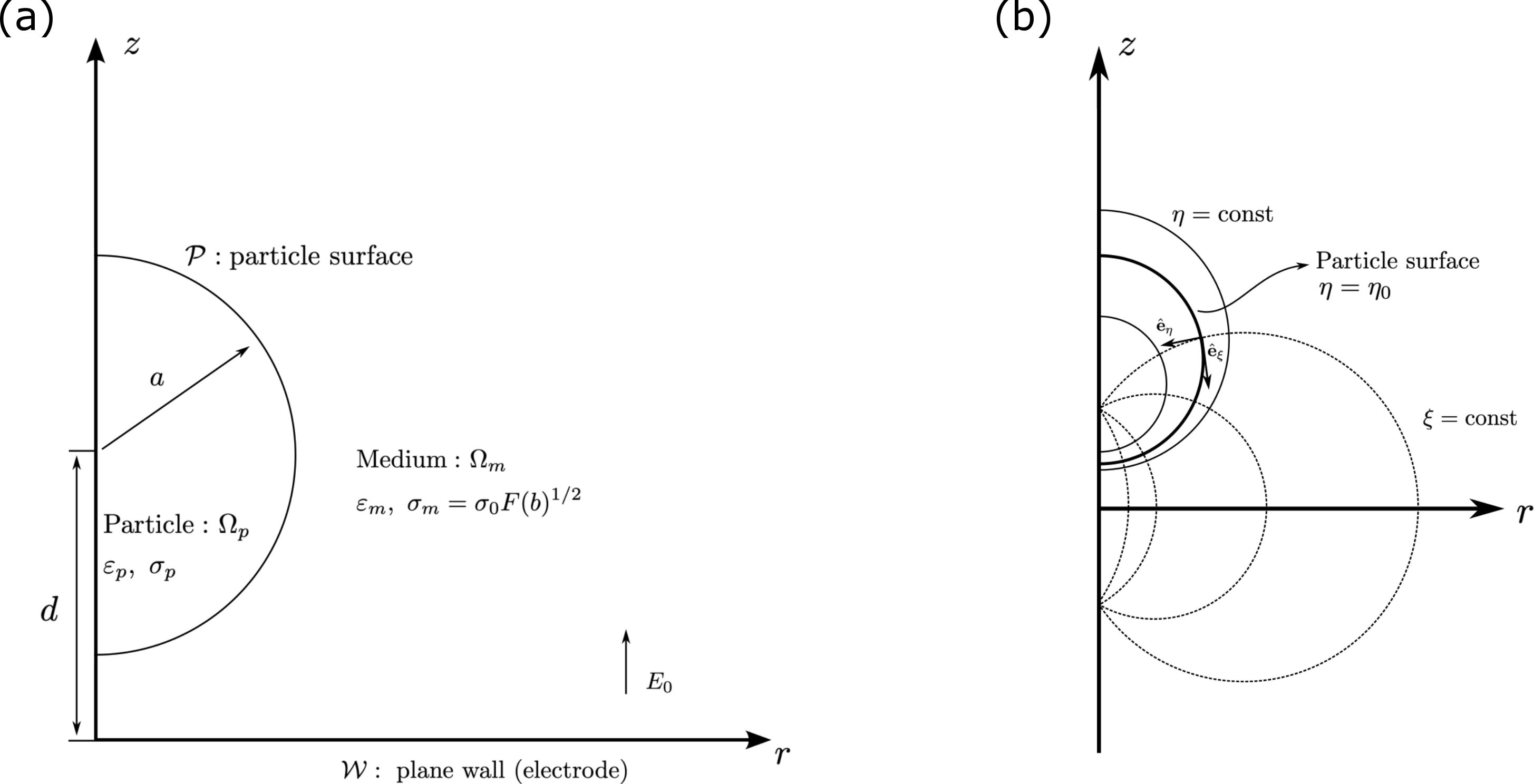}
    \caption{(a) Sketch of the problem in cylindrical coordinates: a spherical particle of radius $a$ centered at $(r,z) = (0,d) $. (b) In bispherical coordinates, the particle surface is given by $\eta=\eta_0=\cosh^{-1}\left(d/a\right)$ and     the electrode surface is specified by $\eta=0$.}
    \label{fig:sketch}
\end{figure}

We consider a spherical particle with radius $a$, permittivity $\eps_p$ and field-independent conductivity $\sigma_p$ placed in a uniform electric field $\bE=E_0\e_z$. The particle center is located at a distance $d$ above a planar electrode at $z = 0$, see \reffig{fig:sketch}a for a sketch of the problem.
We rescale all  variables with particle radius $a$, applied electric field magnitude $E_0$, and the elecrohydrodynamic time scale $t_\ehd = \mu/(\eps_m E_0^2)$.  The dimensionless equations for the electric potential, fluid flow, and charge conservation in the bulk are \citep{Saville:1997}
\begin{eqnarray}
	&& 
	\nabla\cdot(\tilde \sigma \bE)=0\,,\quad \bE=-\nabla\Phi
	 \label{dimensionless_eq1}\\
	&& -\bnabla p + \nabla^2 \bv = -\bE\nabla\cdot\bE,~\bnabla \cdot \bv = 0\,.
	 \label{dimensionless_eq2}
\end{eqnarray}
where the Coulomb force on the fluid is obtained from the Maxwell stress tensor $\bT=\bE\bE-\frac{1}{2}E^2\bI$  as $\nabla\cdot\bT=\bE\nabla\cdot\bE$.
 $\tilde{\sigma}$ denotes the dimensionless  conductivity, $\tilde{\sigma}= \sigma/\sigma_0$. 
The dimensionless boundary conditions are summarized below
\begin{eqnarray}
  &\Phi_m = \Phi_p &\mbox{continuous potential on the particle surface }~\calP, \\
  &\tilde{\sigma}_m \bn \cdot \bE_m = \tilde{\sigma}_p \bn \cdot \bE_p &\mbox{continuous normal current on}~ \calP, \\
  &\Phi_m = 0 &\mbox{grounded electrode on }~\calW,\\
  &\bv = \bs{0} &\mbox{no slip on}~ \calW~\&~\calP, 
  \label{dimensionless_bc4}
\end{eqnarray}
Another dimensionless parameter, the conductivity mismatch $\beta = (\tilde{\sigma}_p-1)/(\tilde{\sigma}_p+1)$, is used in this paper. The extreme cases of perfectly insulating ($\sigma_p=0$) and perfectly conducting ($\sigma_p\rightarrow \infty$)  particles correspond to $\beta=-1$ and $\beta=1$, respectively. Far from the particle, the electric field is undisturbed and uniform, and the electrohydrodynamic flow vanishes.

To solve the problem, we find it convenient to use
bispherical coordinates $(\xi,\eta,\varphi)$ (see \reffig{fig:sketch}b), which are related to the cylindrical coordinates $(r,\varphi,z)$ as follows
\begin{equation} \label{bispherical_def}
	r = \frac{c}{h}\sin\xi,~z = \frac{c}{h}\sinh\eta, h \equiv \cosh\eta-\cos\xi,
\end{equation}
where $c$ is a geometric constant related to the gap between the spherical particle and the electrode, $c = \sqrt{(2+\delta)\delta}$, with $\delta=(d-a)/a$. In bispherical coordinates, the particle surface and the electrode  are iso-surfaces of the coordinate $\eta$,
\begin{equation}
	\calP:~\eta = \eta_0 = \cosh^{-1}(1+\delta),\quad \calW:~\eta = 0.
\end{equation}

\section{Solution  \label{sec:Solution}}
In general,  \refeq{dimensionless_eq1}-\refeq{dimensionless_eq2} can only be solved numerically. Analytical progress can be made assuming small changes in the fluid conductivity with local field intensity. In this case, we develop an asymptotic analysis based on the linearization of 
\refeq{onsager}
for  $b\ll 1$
\begin{equation}
\label{linearS}
\sigma_m=\sigma_0\left(1+\gamma E\right)\,, \quad  \gamma=\frac{e^3}{16 \pi \eps_m (k_BT)^2}\,.
\end{equation}
The dimensionless medium conductivity as a function of dimensionless electric field strength becomes $\tilde{\sigma}_m = 1 + \epsilon E$, where $\epsilon = \gamma E_0 $ is a dimensionless parameter quantifying  the magnitude of the conductivity change by the electric field. The small conductivity variation assumption implies $\epsilon \ll 1$. Estimating $\gamma$ shows that its order of magnitude is $\sim 10^{-6} {\rm m/V}$, hence  the linear approximation is valid for applied electric field less than  $1 {\rm MV/m}$, typical for the experiments \citep{zhang2021quincke, Gerardo:2019}.
$\epsilon \ll 1$ allows for an analytical solution in terms of a regular perturbation series
\begin{eqnarray}
	&&\Phi = \Phi^{\0} + \epsilon \Phi^{(1)}, \label{expansion1} \\
	&&\bE = -\bnabla \Phi = \bE^{(0)} + \epsilon\bE^{(1)}, \\
	&&~p = \epsilon p^{(1)},~\bu = \epsilon\bu^{(1)}. \label{expansion3}
\end{eqnarray}
The leading order problem (with superscripts $(0)$) corresponds to a spherical particle suspended in a charge-free fluid with constant, field-independent conductivity given by $\sigma_0$, which is exactly the leaky dielectric model. The solution predicts no flow about a solid particle  and attraction to a nearby electrode \citep{wang2022particle}, with a force decaying in the far field  as $1/d^4$, where $d$ is the distance between the colloid center and the electrode. The solution is summarized in Appendix \ref{LO}. 

\subsection{The electrohydrodynamic flow}
Here we analyze the correction due to field-dependent conductivity  to the leading-order solution obtained from  the leaky dielectric model. Substituting the linearized relation \refeq{linearS} and expansions \refeq{expansion1}-\refeq{expansion3} into \refeq{dimensionless_eq1}-\refeq{dimensionless_bc4} and collecting terms at $O(\epsilon)$, we find the flow in the suspending fluid  satisfies the following equations,
\begin{eqnarray}
& \left. 
\begin{array}{c}
\displaystyle
  -\bnabla p^{(1)} + \nabla^2 \bu^{(1)} = -\bff \\
\displaystyle
\bnabla\cdot \bu^{(1)} = 0
\end{array} \right\}
&\mbox{in}~\Omega_m \label{EpsEq1}, \\
& \bu^{(1)} = \bs{0} &\mbox{on}~\calW~\&~\calP, \label{EpsBC1}
\end{eqnarray}
where $\bff$ is the Coulomb force on the fluid,$\bff = \left(\nabla\cdot \bE^\1\right) \bE^\0$. Note that by \refeq{rhof}, $\bff$ can be directly related to the leading order ($O(\epsilon)$) solution, $\nabla\cdot\bE^\1 = - \bE^\0 \cdot \bnabla E^\0 $. 
We find the solution {of \refeq{EpsEq1}-\refeq{EpsBC1}} as a superposition of a particular solution $(p^\sfP,\bv^\sfP)$ and a homogeneous solution $(p^\sfH,\bv^\sfH)$,
\begin{equation}
	p^{(1)} = p^\sfP + p^\sfH,~\bu^{(1)} = \bu^\sfP + \bu^\sfH.
\end{equation}
The particular solution solves the nonhomogeneous Stokes equation in a particle-free space $\Omega_p\cup\Omega_m$, which is the upper half space $z > 0$, with non-slip boundary condition on the electrode $\calW$,
\begin{eqnarray}
& \left. 
\begin{array}{c}
\displaystyle
  -\bnabla p^\sfP + \nabla^2 \bu^\sfP = -\tilde{\bff} \\
\displaystyle
\bs{\nabla}\cdot \bu^\sfP = 0
\end{array} \right\}
&\mbox{in}~\Omega_p\cup\Omega_m, \label{ExtensionEq} \\
& \bu^\sfP = \bs{0} &\mbox{on}~\calW. \label{ExtensionBC}
\end{eqnarray}
Here $\tilde{\bff}$ is the extended force which is equal to the Coulomb force $\bff$ in the medium phase $\Omega_m$ and zero in the particle phase $\Omega_p$,
\begin{equation}
	\tilde{\bff} =
	\begin{cases}
		\bff & \mbox{in}~ \Omega_m, \\
		\bs{0} & \mbox{in}~ \Omega_p.
	\end{cases}
\end{equation}
To compensate for the non-zero velocity at the particle surface from the particular solution, the homogeneous solution is added, which solves the  Stokes equations with velocity $-\bv^\sfP$ at the particle surface and no-slip condition on the electrode,
\begin{eqnarray}
& \left. 
\begin{array}{c}
\displaystyle
-\bnabla p^\sfH + \nabla^2 \bu^\sfH = 0 \\
\displaystyle
\bs{\nabla}\cdot \bu^\sfH = 0
\end{array} \right\}
&\mbox{in}~\Omega_m, \\
& \bu^\sfH = -\bv^\sfP &\mbox{on}~\calP, \\
& \bu^\sfH = 0 &\mbox{on}~\calW.
\end{eqnarray}

\subsubsection{Particular solution}
The particular solution solves the incompressible Stokes equation with extended force $\tilde{\bff}$ in the upper half space $z > 0$. In cylindrical coordinates $(r,\varphi,z)$, \refeq{ExtensionEq} and \refeq{ExtensionBC} read
\begin{eqnarray}
	&&-\frac{\partial p^\sfP}{\partial r} + \left(\scrL_{-1} + \frac{\partial^2}{\partial z^2}\right) u_r^\sfP = -\tilde{f}_r, \label{Extension1}\\
	&&-\frac{\partial p^\sfP}{\partial z} + \left(\scrL_0 + \frac{\partial^2}{\partial z^2}\right) u_z^\sfP = -\tilde{f}_z, \label{Extension2}\\
	&& \frac{1}{r}\frac{\partial}{\partial r}\left(ru_r^\sfP\right) + \frac{\partial u_z^\sfP}{\partial z} = 0, \label{Extension3} \\
	&& u_r^\sfP|_{z=0} = u_z^\sfP|_{z=0}= 0, \label{Extension4}
\end{eqnarray}
where the operator $\scrL_{-n}$ is
\begin{equation}
	\scrL_{-n} = \frac{\partial^2}{\partial r^2} + \frac{1}{r}\frac{\partial}{\partial r} - \frac{n^2}{r^2}. \nonumber
\end{equation}
The problem is solved by applying a Hankel-Fourier transform. The details of the calculation are in Appendix \ref{sec:StokesCylindrical}. The obtained velocity components are
\begin{eqnarray}
	u_r^\sfP(r,z) =&& -\frac{2}{\pi} \int_{0}^{\infty}\int_{0}^{\infty} \frac{\omega\hat{R}(k,\omega)}{(k^2+\omega^2)^2} \left[\cos(\omega z) - (1-kz) e^{-kz}\right] J_1(kr) d\omega dk, \label{vrE} \\
	u_z^\sfP(r,z) =&& \frac{2}{\pi} \int_{0}^{\infty}\int_{0}^{\infty} \frac{k\hat{R}(k,\omega)}{(k^2+\omega^2)^2} \left[\sin(\omega z) - \omega z e^{-kz}\right] J_0(kr) d\omega dk, \label{vzE}
\end{eqnarray}
where $J_0$ and $J_1$ are the Bessel functions of the first kind. $\hat{R}(k,\omega)$ is the transform of the extended Coulomb force $\tilde{\bff}$,
\begin{eqnarray}
 	\hat{R}(k,\omega) = \int_0^\infty \int_0^\infty  \left(k^2\tilde{f}_z J_0(kr)\sin(\omega z) 
	- k\omega  \tilde{f}_r J_1(kr)\cos(\omega z) \right)r drdz. 
 \label{Rhat}
\end{eqnarray}

\subsubsection{Homogeneous solution}
The homogeneous problem is found in bispherical coordinates using the general solution developed by \cite{lee1980motion}. The detailed calculation is presented in Appendix \ref{sec:StokesBispherical}.
The obtained velocity components in the cylindrical coordinate system, $u_r^\sfH$ and $u_z^\sfH$, are in the following form,
\begin{eqnarray}
	u_r^\sfH = && \frac{\sin\xi}{2\sqrt{h}} \sum_{n=0}^{\infty} \left[A_n\sinh(\lambda_n\eta) + B_n\cosh(\lambda_n\eta)\right] P_n(\cos\xi) \nonumber\\
	&& + \sqrt{h} \sum_{n=1}^{\infty} \left[E_n\sinh(\lambda_n\eta) + F_n\cosh(\lambda_n\eta)\right] P_n^1(\cos\xi), \\
	u_z^\sfH = && 	\frac{\sinh\eta}{2\sqrt{h}} \sum_{n=0}^{\infty} \left[A_n\sinh(\lambda_n\eta) + B_n\cosh(\lambda_n\eta)\right] P_n(\cos\xi) \nonumber \\
	&& + \sqrt{h} \sum_{n=0}^{\infty} \left[C_n\sinh(\lambda_n\eta) + D_n\cosh(\lambda_n\eta)\right] P_n(\cos\xi),
\end{eqnarray}
where $\lambda_n = n+1/2$, $P_n$ are the Legendre polynomials and $P_n^1$ are the associated Legendre polynomials. The procedure to obtain the  coefficients $A_n,\, B_n, \,C_n,\,D_n, \,E_n, \,F_n$ is given in Appendix \ref{sec:StokesBispherical}.

\subsection{Electrohydrodynamic force on the particle near the electrode\label{sec:Force}}

The electrohydrodynamic force on the particle is conveniently calculated using  the Lorentz reciprocal theorem,
\begin{equation} \label{Reciprocal}
	\oint_{\partial \Omega_m} \bu^{(1)} \cdot ( \boldsymbol{\bsigma}' \cdot \bn ) dS - \int_{\Omega_m} \bu^{(1)} \cdot (\bnabla\cdot\bsigma') dV = \oint_{\partial \Omega_m} \bu' \cdot ( \bsigma^{(1)} \cdot \bn ) dS - \int_{\Omega_m} \bu' \cdot (\bnabla\cdot\bsigma^{(1)}) dV,
\end{equation}
The boundary of the medium phase is $\partial \Omega_m = \calP \cup \calW \cup \calS_\infty$, where $\calS_\infty$ is a surface far from the particle. Due to the axial symmetry, the force has only a $z$-component. Note that in \refeq{Reciprocal} the normal $\bn$ on the particle surface $\calP$ points into the particle phase. 

Te velocity and stress fields $\bv'$ and $\bsigma'$ are the solution of the problem for a translating sphere near a planar wall,
\begin{eqnarray}
& \left. 
\begin{array}{c}
\displaystyle
-\bnabla p' + \nabla^2 \bu' = 0 \\
\displaystyle
\bnabla\cdot\bu' = 0
\end{array} \right\}
&\mbox{in}~\Omega_m, \nonumber \\
& \bu' = \bs{0} &\mbox{on}~\calW, \nonumber \\
& \bu' = \e_z &\mbox{on}~\calP, \nonumber
\end{eqnarray}
where $\e_z$ is the unit vector in $z$-direction. The equations are solved in bispherical coordinates, see Appendix \ref{sec:StokesBispherical}.

The particle has zero net charge. The leading order electric field strength has the following far-field behavior $E^{(0)} = 1 + O(\rho^{-3})$, where $\rho = \sqrt{r^2 + z^2}$. Consequently, we have $\|\bff\| \sim O(\rho^{-4})$ as $\rho \to \infty$. Therefore, in the far-field, the velocity $\bu^{(1)}$ and stress $\bsigma^{(1)}$ decay as $O(\rho^{-2})$ and  $O(\rho^{-3})$, respectively. From \cite{blake1974fundamental}, the velocity $\bu'$ and stress $\bsigma'$ decay as $O(\rho^{-3})$ and $O(\rho^{-4})$. Consequently, integrals over the infinite surface $\calS_\infty$ on both sides of the reciprocal identity \refeq{Reciprocal} vanish,
\begin{equation} \label{inf_integral}
	\int_{\calS_\infty} \bu^{(1)} \cdot (\bsigma'\cdot\bn) dS = \int_{\calS_\infty} \bu' \cdot (\bsigma^{(1)}\cdot\bn) dS = 0.
\end{equation}
Thus, the dimensionless electrohydrodynamic force on the particle, $C_f$, is calculated from the following volume integral,
\begin{equation} \label{Cf}
	C_f = -\int_{\calP} \e_z \cdot (\bsigma^{(1)} \cdot \bn) dS = \int_{\Omega_m} \bu' \cdot \bff dV,
\end{equation}
The dimensional form of the hydrodynamics force is $F = \gamma \varepsilon_m a^2|E_0^3| C_f$, where the absolute value indicates that the direction of the force is independent of the direction of the applied electric field. 

The  volume integral \refeq{Cf}  is computed in bispherical coordinates, which conveniently map the medium phase $\Omega_m$ onto a bounded rectangle region
\begin{equation}
	C_f = 2\pi c^3 \int_0^\pi d\xi \int_0^{\eta_0} d\eta \bu'\cdot \bff \frac{\sin\xi}{h^3}. \nonumber
\end{equation}
This double integral is evaluated numerically with Gauss quadratures. Using the volume integral to find the force coefficient has the following advantages. First, it does not require the solution of the tractions due to flow field $\bu^\1$, which involves numerical evaluation of the integral transforms \refeq{vrE}-\refeq{Rhat}. Second, the velocity field $\bu'$ in the volume integral has an analytical solution in bispherical coordinates. The Coulomb force is evaluated by differentiating the leading order electric field, which also has an analytical solution in bispherical coordinates.


\section{Results and discussion \label{sec:Results}}
\begin{figure}
    \centering
    \includegraphics[width=\linewidth]{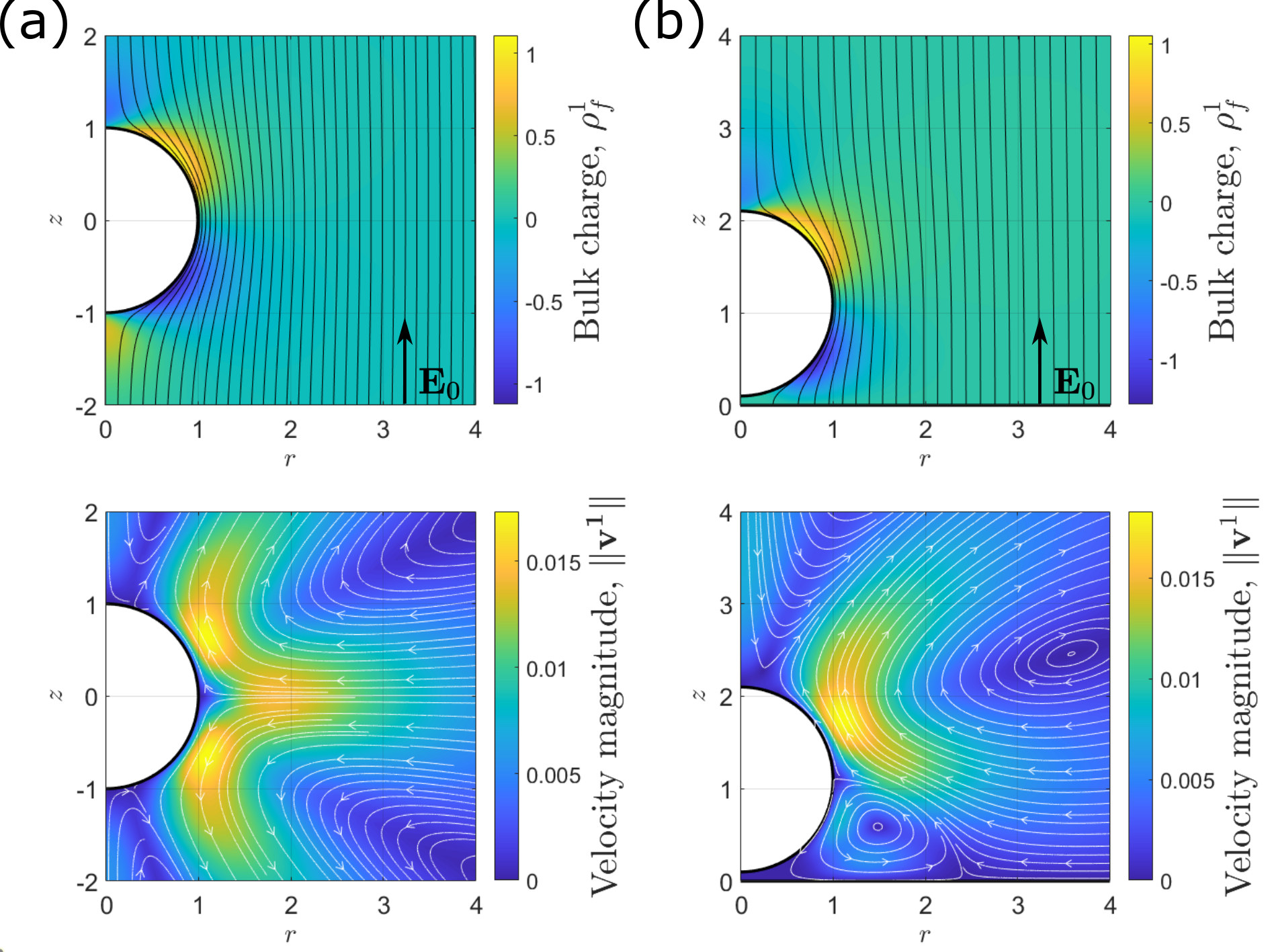}
    \caption{Electric field lines and flow streamlines about an insulating sphere ($\beta = -1$) in (a) unbounded domain, and (b) near the electrode $\delta = 0.1$. The color map in the plots for the electric field shows the magnitude of the induced charge. The color map in the plots for the flow shows the magnitude of the velocity field.}
    \label{fig:Insulation_fields}
\end{figure}

\begin{figure}
    \centering
    \includegraphics[width=\linewidth]{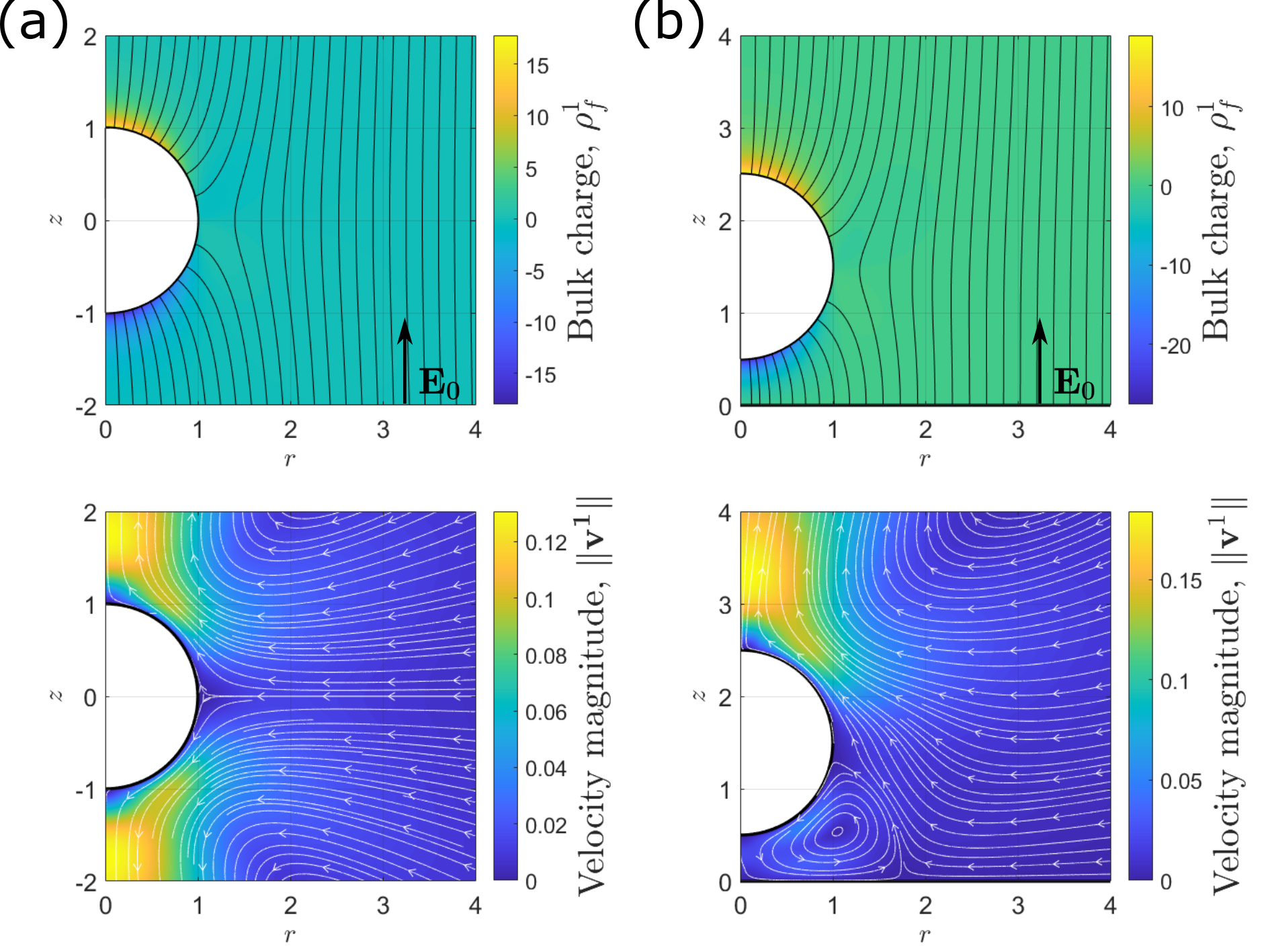}
    \caption{Electric field lines and flow streamlines about a conducting ($\beta = 1$) particle (a) unbounded domain, and (b) near the electrode $\delta = 0.5$. The color map in the plots for the electric field shows the magnitude of the induced charge. The color map in the plots for the flow shows the magnitude of the velocity field.}
    \label{fig:Conducting_fields}
\end{figure}
The leading order electric field, the induced bulk charge, and the resulting $O(\epsilon)$ flow about conducting and insulating particles are shown in the unbounded domain in \reffig{fig:Insulation_fields}a, \reffig{fig:Conducting_fields}a, and for a particle 
close to the electrode in \reffig{fig:Insulation_fields}b, \reffig{fig:Conducting_fields}b.
The unbounded problem with the particle centered at the origin is solved numerically in spherical coordinates with the Chebyshev collocation method (see Appendix \ref{sec:unbounded}); it agrees with the bounded solution in the limit where the distance to the electrode is large. In all cases, the bulk charge is localized near the particle surface, where the electric field nonuniformities and resulting conductivity gradients are the largest. However, the charge distribution and fluid flow depend strongly on the particle conductivity. 

In the case of the insulating particle, the charge distribution resembles an octupole. The positive bulk charge above the equator and below the pole facing the electrode drives upward flows, while the negative charge drives flow in the opposite direction. The flows converge and lead to a total of five dividing streamlines (axisymmetric surfaces in 3D): fluid is drawn towards the particle at the equator and the poles, and pushed out  in between.
The bulk charges are mostly localized above and below the equator. Correspondingly, the velocity magnitude peaks near the equator. 

In the case of a conducting particle, the bulk charge distribution presents a dipole pattern, i.e., opposite-sign charges localize near the poles. The flow is driven away from the poles, leading to inflow toward the equator.

The electrohydrodynamic flows about the conducting and the insulating particles differ quantitatively and qualitatively. Quantitatively, the bulk charge density and velocity magnitude are greater for the conducting than insulating particle by an order of magnitude. Qualitatively, the recirculation and dividing streamlines near the two poles found in the insulating particle disappear for the conducting particle. Intriguingly, the flow pattern in the conducting particle case resembles the quadrupolar induced-charge electroosmosis but the direction is reversed.

The presence of a planar boundary (the electrode) does not significantly change the electrostatics or the hydrodynamics above the particle. The flow pattern in both insulating and conducting cases is similar to the unbounded results. However, the flows below the particle are geometrically frustrated, and vortices arise near the particle-wall gap. The vortices in both cases are in the counterclockwise direction.

\begin{figure}
	\centerline{\includegraphics[width=\linewidth]{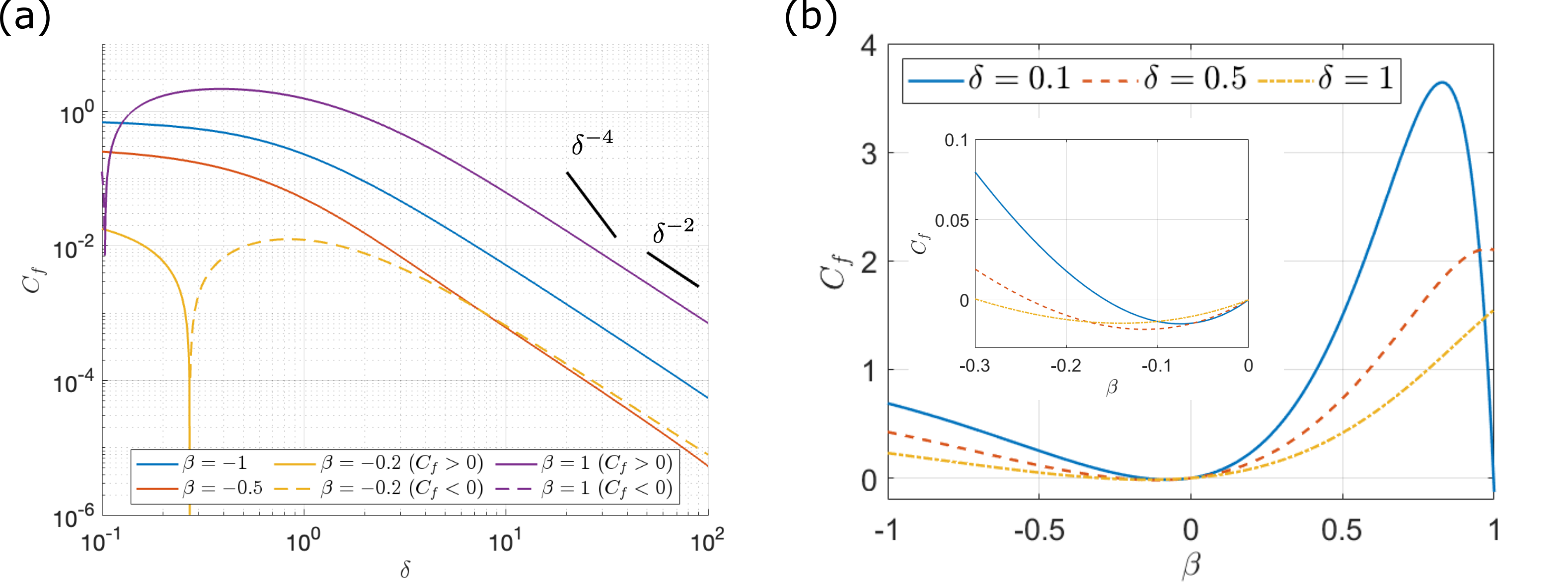}}  
	\caption{(a) The absolute value of force coefficient $|C_f|$ as a function of the dimensionless separation from the electrode, $\delta$, for various particle conductivities, $\beta$. (b). The force coefficient $C_f$ as a function of conductivity mismatch $\beta$ for various $\delta$.}
	\label{fig:Cf}
\end{figure}

The electrohydrodynamic force calculated from \refeq{Cf} is shown in \reffig{fig:Cf}.
\begin{figure}
	\centerline{\includegraphics[width=\linewidth]{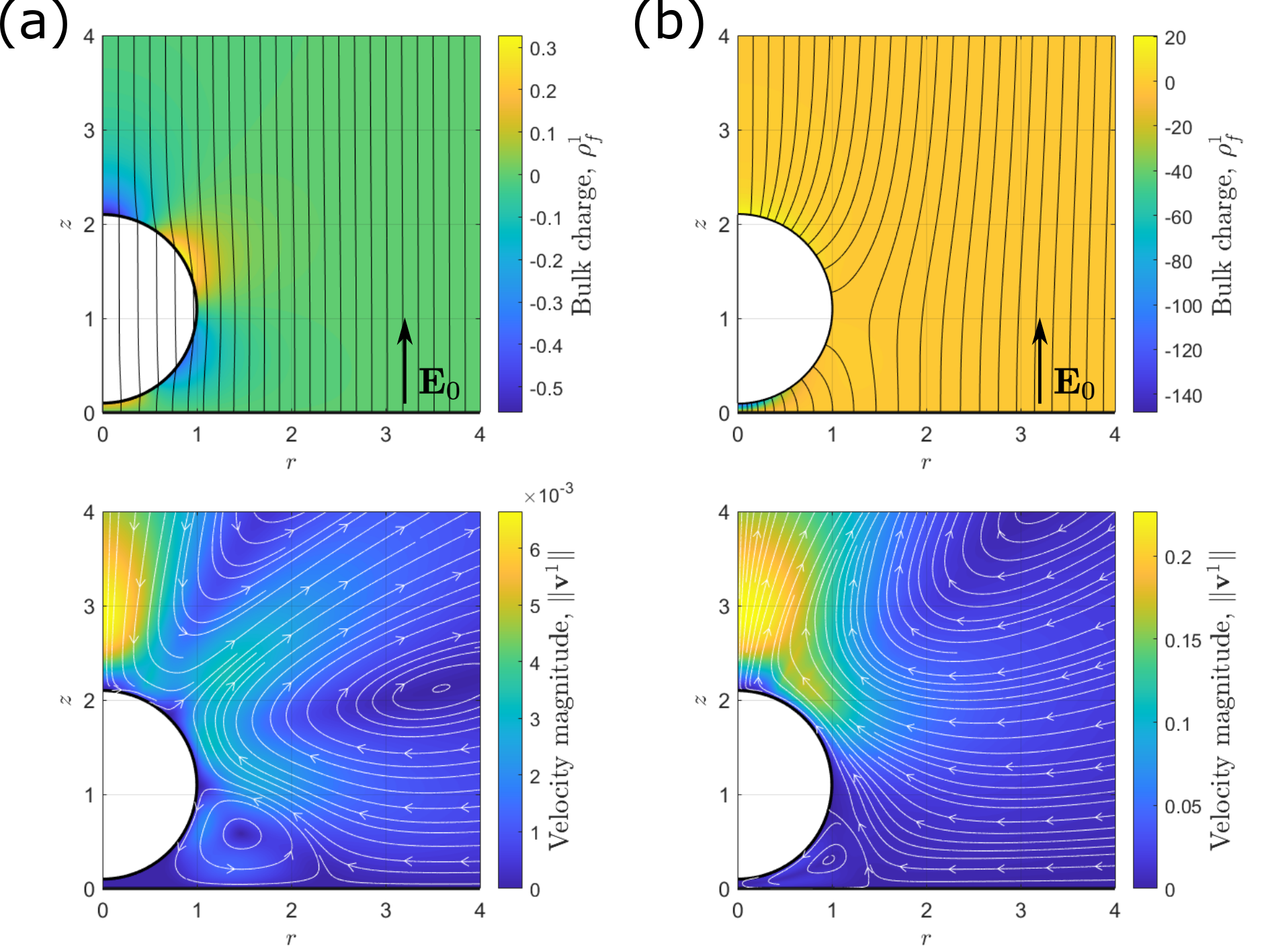}}  
	\caption{Electric field lines and flow streamlines in the case of attractive electrohydrodynamic force. (a) $\delta = 0.1,~\beta = -0.2$ (particle less conducting than the suspending fluid). (b) $\delta = 0.1,~\beta = 1$ (perfectly conducting particle).}
	\label{fig:Attraction}
\end{figure}
\reffig{fig:Cf}a illustrates the effect of particle-electrode separation.
The force magnitude decreases as the particle moves away from the electrode. \reffig{fig:Cf}b shows that the force is also weakened as the difference in the conductivity between the two phases decreases. In general, the force is repulsive (positive). However, the force can be attractive (negative) when $\beta$ is close to 0 and 1. The change of the sign can be seen from dips in \reffig{fig:Cf}a in the cases of $\beta =-0.2$ and $1$. 
The electrostatic field and flow streamlines in these two attracting cases are shown in \reffig{fig:Attraction}. In the first case, the particle is less conducting, and the charge distribution in \reffig{fig:Attraction}a shows an octupole pattern. However, the bulk charge density peaks near the pole far from the electrode, and the corresponding downward flow is dominating, leading to the negative electrohydrodynamic force on the particle (attractive to the electrode). For a conducting particle, the charge distribution, and consequently the Coulomb force, in the thin gap is singularly large. The velocity magnitude in the gap is small. From the balance between the pressure gradient and Coulomb force, $\bnabla p^{(1)} \sim \bff$, we find the pressure drop across the thin gap between the electrode and the particle surface is enormous. The low pressure in the gap region leads to the downward electrohydrodynamic force on the conducting particle. This effect disappears when the gap is large since large gaps allow the fluid to recirculate (see \reffig{fig:Conducting_fields}b) and reduce the singularity of the pressure. However, the high bulk charge density in the thin gap may indicate the violation of our assumption of the small change of conductivity. A more detailed analysis is required in this case.

It is observed in \reffig{fig:Cf}a that the electrohydrodynamic force on the particle decays as $1/d^2$ when the particle is far from the electrode. The flow around a force-free and torque-free particle behaves like a stresslet in the far field. Accordingly, it is expected that the particle migration at large distances from the wall is driven by the flow due to image stresslet,  which decays as inverse square of the distance.

\section{Conclusions}

Nonlinear electrohydrodynamic flows driven by an electric field acting on its own induced space charge about colloidal particles in aqueous electrolyte solutions are well-documented and extensively studied topics. The prime example is the induced-charge electroosmosis   due to the electric field acting on charge accumulated near polarizable surfaces \citep{squires2004induced, ristenpart2004assembly,ristenpart2007electrohydrodynamic}

Here we show that nonlinear electrohydrodynamic flows can arise  about particles suspended in non-polar, leaky dielectric fluids. The flows are driven by space charge generated by spatially varying conductivity due to the Onsager effect.  For a spherical particle in an applied uniform electric field, the resulting flow pattern strongly depends on the particle conductivity. Intriguingly, for a conducting particle the flow pattern is quadrupolar, resembling induced-charge electroosmosis, however, the direction is reversed.  The electrohydrodynamic flow gives rise to a force on the particle, which is in general repulsive. It decays more slowly with the distance to the electrode than the dielectrophoretic attractive force: quadratic vs fourth-power.

The leading order problem in our analysis corresponds to the leaky dielectric model (LDM) of Melcher and Taylor. For a solid particle, there is no flow within the LDM framework and hence the Onsager-effect flow becomes important. However,  in the case of a drop, the LDM flow due to the interfacial charge at the fluid/fluid interface would dominate over the Onsager-effect flow correction.

Our results have direct relevance to particle manipulation and assembly in non-polar fluids. The existence of a repulsive electrohydrodynamic force may explain the reported levitation above the electrode of colloids suspended in hexadecane  \citep{Gerardo:2019}. The electrohydrodynamic flows may also have strong effects on the collective dynamics of the Quincke rollers, a popular model of active matter system \citep{Bartolo:2013, Driscoll:2019}. The present analysis only considers DC electric field. In AC fields, the large disparity in ions mobility may give rise a steady component \citep{Hashemi:2018} and much richer electrohydrodynamic flows and colloidal dynamics.

\section*{Acknowledgements}

This work was partially supported by NSF awards DMS-2108502 and CBET-2126498. 

\appendix

\section{Leading order solution: electric field about a particle near an electrode}\label{LO}
The leading order problem is formulated below and solved in bispherical coordinates defined in \refeq{bispherical_def} \citep{ wang2022particle} (the more general case of a spherical particle between two electrodes is solved in \cite{Wang:2023}). The governing equations are 
\begin{eqnarray}
	& \nabla^2\Phi^{(0)} = 0 &{\rm in}~\Omega_m\cup\Omega_p, \\
	& \Phi_m^{(0)} = \Phi_p^{(0)}  &{\rm on}~\calP, \\
	& \e_\eta \cdot \bE_m^{(0)} = \tilde{\sigma}_p \e_\eta \cdot \bE_p^{(0)} &{\rm on}~ \calP, \\
	& \Phi_m^{(0)} \sim -z &{\rm as}~\sqrt{r^2 + z^2} \to \infty,
\end{eqnarray}
where $\e_\eta$ is the unit vector in $\eta$-direction {and inward normal on particle surface}. In bispherical coordinates, the electric potential is written as follows
\begin{eqnarray}
	\Phi^{(0)}_p =&& -z + \sqrt{h} \sum_{n=0}^{\infty} \tilde{X}_n e^{-\lambda_n\eta} P_n(\cos\xi), \nonumber\\
	\Phi^{(0)}_m =&& -z + 2 \sqrt{h} \sum_{n=0}^{\infty} X_n \sinh(\lambda_n\eta) P_n(\cos\xi), \nonumber
\end{eqnarray}
where $\lambda_n = n+1/2$, $\tilde{X}_n = X_n \left(e^{2\lambda_n\eta_0}-1\right)$, and $P_n$ are the Legendre polynomials. Coefficients $X_n$ are solved from the following tridiagonal system
\begin{equation} \label{SystemEl}
	\calL^e_{n,1} X_{n-1} + \calL^e_{n,2} X_n + \calL^e_{n,3} X_{n+1} = \calR^e_n ~(n \geq 0)
\end{equation}
where $\calL^e_{n,1}$, $\calL^e_{n,2}$, $\calL^e_{n,3}$, and $\calR^e_n$ are
\begin{eqnarray}
	&\calL^e_{n,1} &= n\left( e^{\lambda_{n-1}\eta_0} - \beta e^{-\lambda_{n-1}\eta_0} \right), \nonumber \\
	&\calL^e_{n,2} &= \left(\beta \sinh\eta_0-2\lambda_n\cosh\eta_0\right) e^{\lambda_n\eta_0} - \beta \left(\sinh\eta_0-2\lambda_n\cosh\eta_0\right) e^{-\lambda_n\eta_0}, \nonumber\\
	&\calL^e_{n,3} &= (n+1)\left( e^{\lambda_{n+1}\eta_0} - \beta e^{-\lambda_{n+1}\eta_0} \right), \nonumber\\
	&\calR^e_n &= 2\sqrt{2}\beta \left[ 2\lambda_n e^{-\lambda_n\eta_0} - \cosh\eta_0 \left(n e^{-\lambda_{n-1}\eta_0} + (n+1) e^{-\lambda_{n+1}\eta_0}\right)   \right],  \nonumber
\end{eqnarray}
where $\beta$ is the conductivity mismatch, $\beta = (\tilde{\sigma}_p-1)/(\tilde{\sigma}_p+1)$. The equation of $n=0$ in the system \refeq{SystemEl} only has two terms with $X_0$ and $X_1$ since $\calL^e_{0,1} = 0$. The cylindrical components of electric field strength, $\bE^{(0)} = -\bnabla\Phi^{(0)}$, are given below,
\begin{eqnarray}
	E_{p,r}^{(0)} && = -\frac{\sqrt{h}}{2c} \sum_{n=1}^{\infty} \left( \tilde{X}_{n-1} - 2\tilde{X}_n + \tilde{X}_{n+1}\right) e^{-\lambda_n\eta} P_n^1(\cos\xi), \nonumber \\
	E_{p,z}^{(0)} && = 1-\frac{\sqrt{h}}{2c} \sum_{n=0}^{\infty} \left[n\tilde{X}_{n-1} - (2n+1)\tilde{X}_n + (n+1)\tilde{X}_{n+1}\right] e^{-\lambda_n\eta} P_n(\cos\xi) \nonumber \\
	E_{m,r}^{(0)} && = -\frac{\sqrt{h}}{c} \sum_{n=1}^{\infty} (X_{n-1} - 2X_n + X_{n+1}) \sinh(\lambda_n\eta) P_n^1(\cos\xi) \nonumber\\
	E_{m,z}^{(0)} && = 1 + \frac{\sqrt{h}}{c} \sum_{n=0}^{\infty} [nX_{n-1} - (2n+1)X_n + (n+1)X_{n+1}] \cosh(\lambda_n\eta) P_n(\cos\xi). \nonumber
\end{eqnarray}

\section{Particular solution in cylindrical coordinates \label{sec:StokesCylindrical}}
In this section, we solve equations \refeq{Extension1}-\refeq{Extension4} using integral transforms. The Hankel transform and its inverse are defined below
\begin{eqnarray}
	&&G(k) = \scrH_n[g(r)] = \int_0^\infty rg(r) J_n(kr) dr, \\
	&&g(r) = \scrH^{-1}_n[G(k)] = \int_0^\infty kG(k) J_n(kr) dk,
\end{eqnarray}
where $J_n$ is the Bessel function of the first kind. Hankel transforming the partial differential equations \refeq{Extension1} - \refeq{Extension3} to ordinary differential equations in $z$,
\begin{eqnarray}
	&&kP + \left(\frac{d^2}{d z^2} - k^2 \right) V_r = -F_r, \label{ExtensionHankel1}\\
	&&-\frac{d}{dz} P + \left(\frac{d^2}{d z^2} - k^2\right) V_z = -F_z, \label{ExtensionHankel2} \\
	&&kV_r + \frac{d}{dz}V_z = 0. \label{ExtensionHankel3}
\end{eqnarray}
The Hankel transform variables are defined below
\begin{eqnarray}
	&&P(z,k),V_z(z,k),F_z(z,k) = \scrH_0[p^\sfP(r,z),v_z^\sfP(r,z),\tilde{f}_z(r,z)], \\
	&&V_r(z,k),F_r(z,k) = \scrH_1[v_r^\sfP(r,z),\tilde{f}_r(r,z)].
\end{eqnarray}
Note that $F_r(z,k)$ and $F_z(z,k)$ are continuous even though the extended force $\tilde{\bff}$ has a finite jump on the particle surface $\calP$. These ordinary differential equations are simplified to a single equation in terms of $V_z$,
\begin{equation} \label{EqnVz}
	V''''_z - 2k^2 V''_z + k^4 V_z = R, R = k^2 F_z + k F'_r,
\end{equation}
where the prime stands for the derivative with respect to $z$. From the non-slip boundary condition on the plane electrode, we find the boundary conditions at $z = 0$ are $V_z|_{z=0} = V_r|_{z=0} = 0$, which is equivalent to $V_z|_{z=0} = V'_z|_{z=0} = 0$. The far-field boundary conditions are $V_z,~V'_z \to 0$ as $z \to +\infty$.

We write the solution as the superposition of the particular solution $U(z,k)$ and homogeneous solution $W(z,k)$, $V_z(z,k) = U(z,k)+W(z,k)$. The particular solution is solved from the following equation and boundary conditions
\begin{equation} \label{VzParticular}
	U'''' - 2k^2 U'' + k^4 U = R,~U|_{z=0} = U''|_{z=0} = 0.
\end{equation}
The physical interpretation of the boundary condition $U''|_{z = 0} = 0$ is that the shear stress on the plane wall is zero. Inverse Hankel transforming the particular solution $U$ yields the flow driven by force $\tilde{\bff}$ with zero shear stress and permeability on $\calW$. We prescribe the second order derivative to be zero so that $U$ could be constructed by Fourier-sine transform,
\begin{eqnarray}
	&&\hat{G}(\omega) = \scrF_s[G(z)] = \int_0^\infty G(z) \sin(\omega z) dz \\
	&&G(z) = \scrF_s^{-1}[\hat{G}(\omega)] = \frac{2}{\pi} \int_0^\infty \hat{G}(\omega) \sin(\omega z) d\omega
\end{eqnarray}
Fourier-sine transforming \refeq{VzParticular} yields
\begin{equation}
	\hat{U}(k,\omega) = \frac{\hat{R}(k,\omega)}{(\omega^2 + k^2)^2},
\end{equation}
where $\hat{U}(k,\omega) = \scrF_s[U(z,k)]$ and $\hat{R}(k,\omega)$ is given in \refeq{Rhat}. Inverse transforming $\hat{U}(k,\omega)$ gives
$$
U(z,k) = \frac{2}{\pi} \int_{0}^{\infty} \frac{\hat{R}(k,\omega)}{(\omega^2 + k^2)^2} \sin(\omega z)d\omega.
$$
The homogeneous solution $W(z,k)$ is solved from the following equation and boundary conditions,
\begin{equation}
	W'''' - 2k^2 W'' + k^4 W= 0,~W|_{z=0} = 0,~W'|_{z=0} = -U'|_{z=0}.
\end{equation}
For convenience, we note
\begin{equation}
	U'|_{z=0} = U_0(k) = \frac{2}{\pi} \int_{0}^{\infty} \frac{\hat{R}(k,\omega)}{(\omega^2+k^2)^2} \omega d\omega. \nonumber
\end{equation}
Solving the homogeneous equation and keeping terms that vanish as $z \to \infty$ and at $z=0$ gives
\begin{equation}
	W(z,k) = - U_0(k) z e^{-kz}. \nonumber
\end{equation}
Adding up $U(z,k)$ and $W(z,k)$ gives the expression of $V_z(z,k)$
\begin{equation}
	V_z(z,k) = \frac{2}{\pi} \int_{0}^{\infty} \frac{\hat{R}(k,\omega)}{(\omega^2 + k^2)^2} \left[\sin(\omega z) - \omega z e^{-kz}\right] d\omega.
	\nonumber
\end{equation}
Substituting into \refeq{ExtensionHankel3} gives the expression for $V_r(z,k)$
\begin{equation}
	V_r(z,k) = -\frac{2}{k\pi}\int_{0}^{\infty} \frac{\omega \hat{R}(k,\omega)}{(\omega^2+k^2)^2} \left[\cos(\omega z) - (1-kz) e^{-kz}\right] d\omega.
	\nonumber
\end{equation}
Inverse Hankel transforming $V_r(z,k)$ and $V_z(z,k)$ gives the solution \refeq{vrE} and \refeq{vzE}.

\section{Homogeneous solution in bispherical coordinates \label{sec:StokesBispherical}}
In this section, we summarize the solution of the homogeneous Stokes equation 
\begin{equation}
	-\bnabla p + \nabla^2 \bv = 0, ~\bnabla \cdot \bv = 0, \nonumber
\end{equation}
for arbitrary velocity distribution on the particle surface $\calP$ and zero velocity on the electrode $\calW$,
\begin{eqnarray}
	&&u_r|_\calW = u_z|_\calW = 0, \\
	&&u_r|_\calP = v_1,\,~u_z|_\calP = v_2.
\end{eqnarray}
For a particle translating in the $z$-direction, $v_2=1$ and $v_1=0$. 
In bispherical coordinates, the asymmetric velocity field could be written as follows \citep{lee1980motion},
\begin{eqnarray}
	&& p = p_0,~u_r = \frac{rp}{2} + u_0,~u_z = \frac{zp}{2} + w_0, \\
	&& p_0 = \frac{\sqrt{h}}{c} \sum_{n=0}^{\infty} \left[A_n\sinh(\lambda_n\eta) + B_n\cosh(\lambda_n\eta)\right] P_n(\cos\xi), \label{p0} \\
	&& u_0 = \sqrt{h} \sum_{n=1}^{\infty} [E_n\sinh(\lambda_n\eta) + F_n\cosh(\lambda_n\eta)] P_n^1(\cos\xi), \label{u0} \\
	&& w_0 = \sqrt{h}\sum_{n=0}^{\infty} [C_n\sinh(\lambda_n\eta) + D_n\cosh(\lambda_n\eta)]P_n(\cos\xi). \label{w0}
\end{eqnarray}
where $\lambda_n = n + 1/2$, $P_n$ are the Legendre polynomials, and $P_n^1$ are the associated Legendre polynomials. In the axisymmetric case, the incompressibility reads
\begin{equation}
	\frac{\partial u_r}{\partial r} + \frac{u_r}{r} + \frac{\partial u_z}{\partial z} = 0.
\end{equation}
From the boundary condition $u_r|_\calW = 0$, we obtain $D_n = 0 ~(n\geq0)$. Substituting $u_r|_\calW = 0$ into the incompressibility, we find
\begin{equation}
	\left. \frac{\partial u_z}{\partial z} \right|_\calW = 0, \nonumber
\end{equation}
which leads to
\begin{equation}
	p|_\calW = -2 \left. \frac{\partial w_0}{\partial z}\right|_\calW. \nonumber
\end{equation}
Substituting the expressions of $p$ and $w_0$ into the identity, we obtain
\begin{equation} \label{Bn0}
	B_n = nC_{n-1} - (2n+1)C_n + (n+1)C_{n+1} ~(n\geq0).
\end{equation}
From the boundary condition $u_z|_\calP = f_2$, we obtain the following equation
\begin{eqnarray}
	\frac{1}{2}\sinh\eta_0 \left[A_n\sinh(\lambda_n\eta_0) + B_n\cosh(\lambda_n\eta_0)\right] + C_n \cosh\eta_0 \sinh(\lambda_n\eta_0) \nonumber \\
	 - \frac{n}{2n-1} \sinh(\lambda_{n-1}\eta_0)C_{n-1} - \frac{n+1}{2n+3} \sinh(\lambda_{n+1}\eta_0)C_{n+1} = \alpha_n,
	 \label{An0Eqn}
\end{eqnarray}
where $\alpha_{n}$ come from the following expansion
\begin{equation}
	\sqrt{\cosh\eta_0 -\cos\xi} v_2 = \sum_{n=0}^{\infty} \alpha_n P_n(\cos\xi). \nonumber
\end{equation}
Expansion coefficients $\alpha_{n}$ are calculated using the orthogonality of Legendre polynomials
\begin{equation} \label{Expansion_Alpha}
	\alpha_n = \lambda_n \int_0^\pi \sqrt{\cosh\eta_0 - \cos\xi} v_2 P_n(\cos\xi)\sin\xi d\xi ~(n \geq 0).
\end{equation}
Substituting \refeq{Bn0} into \refeq{An0Eqn} allows to write $A_n^0$ in terms of $C_n^0$,
\begin{equation} \label{An0}
	A_n = \frac{2 \alpha_n}{\sinh\eta_0\sinh(\lambda_n\eta_0)} - 2 \kappa_n \left(\frac{n}{2n-1}C_{n-1} - C_n + \frac{n+1}{2n+3}C_{n+1}\right) ~(n \geq 0), 
\end{equation}
where coefficients $\kappa_n = \lambda_n\coth(\lambda_n\eta_0) - \coth\eta_0$. The boundary condition $u_r|_\calW = 0$ leads to
\begin{equation}
	u_0|_\calW = -\frac{r}{2}p_0|_\calW = r \left. \frac{\partial w_0}{\partial z}\right|_\calW. \nonumber
\end{equation}
Substituting \refeq{u0} and \refeq{w0} into the equation, we find
\begin{equation} \label{Fn0}
	F_n = \frac{1}{2} (C_{n+1} - C_{n-1}) ~(n\geq1).
\end{equation}
Paralleling the two boundary conditions on $\calP$ gives
\begin{equation}
-\frac{\sin\xi}{\sinh\eta_0} w_0|_\calP + u_0|_\calP = g, ~g = v_1 - \frac{\sin\xi}{\sinh\eta_0} v_2.
\end{equation}
Substituting \refeq{u0} and \refeq{w0} into the equation, we obtain
\begin{eqnarray}
	\frac{1}{2n-1}\sinh(\lambda_{n-1}\eta_0) C_{n-1} - \frac{1}{2n+3}\sinh(\lambda_{n+1}\eta_0) C_{n+1} \nonumber\\
	+ \sinh\eta_0 [E_n\sinh(\lambda_n\eta_0) + F_n\cosh(\lambda_n\eta_0)] = \sinh\eta_0 \beta_n, 
	\label{En0Eqn}
\end{eqnarray}
where $\beta_n$ are coefficients in the following expansion
\begin{equation}
	\frac{g}{\sqrt{\cosh\eta_0 -\cos\xi}} = \sum_{n=1}^{\infty} \beta_n P_n^1(\cos\xi). \nonumber
\end{equation}
From the orthogonality of associated Legendre polynomials, we have
\begin{equation} \label{Expansion_Beta}
	\beta_n = \frac{\lambda_n}{n(n+1)} \int_0^\pi \frac{g}{\sqrt{\cosh\eta_0 - \cos\xi}} P_n^1(\cos\xi) \sin\xi d\xi ~(n \geq 1).
\end{equation}
In the case, $v_1 = 0$ and $v_2 = 1$ (translating spherical particle), the expansion coefficients $\alpha_n$ and $\beta_n$ are calculated analytically
\begin{eqnarray}
	&& \alpha_n = \sqrt{2} e^{-\lambda_n\eta_0} \left[\cosh\eta_0 - \frac{n e^{\eta_0}}{2n-1} - \frac{(n+1) e^{-\eta_0}}{2n+3}\right], \nonumber\\
	&& \beta_n = \frac{2\sqrt{2}}{\sinh\eta_0} e^{-\lambda_n\eta_0} \left[\cosh\eta_0 - \frac{(n-1)e^{\eta_0}}{2n-1} - \frac{(n+2)e^{-\eta_0}}{2n+3}\right]. \nonumber
\end{eqnarray}
Substituting \refeq{Fn0} into \refeq{En0Eqn}, we write coefficients $E_n$ in terms of $C_n$,
\begin{equation} \label{En0}
	E_n = \frac{\beta_n}{\sinh(\lambda_n\eta_0)} + \kappa_n \left(\frac{1}{2n-1}C_{n-1} - \frac{1}{2n+3}C_{n+1}\right) ~(n \geq 1). 
\end{equation}
From the incompressibility, we have the following two equations
\begin{eqnarray}
	-\frac{1}{2}nA_{n-1} + \frac{5}{2}A_n + \frac{1}{2}(n+1)A_{n+1} - (n-1)nE_{n-1} + 2n(n+1)E_n \nonumber\\
	- (n+1)(n+2)E_{n+1} - nD_{n-1} + (2n+1)D_n - (n+1)D_{n+1} = 0, \label{Incompressibility1} \\
	-\frac{1}{2}nB_{n-1} + \frac{5}{2}B_n + \frac{1}{2}(n+1)B_{n+1} - (n-1)nF_{n-1} + 2n(n+1)F_n \nonumber\\
	- (n+1)(n+2)F_{n+1} - nC_{n-1} + (2n+1)C_n - (n+1)C_{n+1} = 0. \label{Incompressibility2}
\end{eqnarray}
It is verified that \refeq{Incompressibility2} is satisfied automatically using \refeq{Bn0} and \refeq{Fn0}. Plugging \refeq{An0} and \refeq{En0} into \refeq{Incompressibility1} yields a linear system for $C_n$,
\begin{equation} \label{Cn0Eqn}
	\calL^h_{n,1} C_{n-1} + \calL^h_{n,2} C_n + \calL^h_{n,3} C_{n+1} = \calR^h_n ~(n \geq 0).
\end{equation}
$\calL^h_{n,1}$, $\calL^h_{n,2}$, $\calL^h_{n,3}$, and $\calR^h_n$ are listed below,
\begin{eqnarray}
	\calL^h_{n,1} =&& -n\kappa_{n-1} + \frac{n(2n-3)}{2n-1}\kappa_n, \nonumber\\
	\calL^h_{n,2} =&& \frac{n(2n-1)}{2n+1}\kappa_{n-1} + 5\kappa_n - \frac{(n+1)(2n+3)}{2n+1}\kappa_{n+1}, \nonumber\\
	\calL^h_{n,3} =&& -\frac{(n+1)(2n+5)}{2n+3}\kappa_n + (n+1)\kappa_{n+1}, \nonumber\\
	\calR^h_n =&& \frac{n}{\sinh(\lambda_{n-1}\eta_0)} \left[\frac{\alpha_{n-1}}{\sinh\eta_0} + (n-1)\beta_{n-1}\right] - \frac{1}{\sinh(\lambda_n\eta_0)} \left[\frac{5\alpha_n}{\sinh\eta_0} + 2n(n+1)\beta_n\right] \nonumber\\
	&& - \frac{n+1}{\sinh(\lambda_{n+1}\eta_0)} \left[\frac{\alpha_{n+1}}{\sinh\eta_0} - (n+2)\beta_{n+1}\right]. \nonumber
\end{eqnarray}
The equation of $n = 0$ in the system \refeq{Cn0Eqn} has two terms involving $C_0$ and $C_1$ since $\calL^h_{0,1} = 0$.

\section{Flow and electric field in an unbounded domain \label{sec:unbounded}}
The unbounded problem is solved numerically in the spherical coordinates $(r,\theta,\varphi)$ originated at the center of the particle. The electric potential solved from the leading order electrostatics is given below,
\begin{equation}
	\Phi^0_p = -\frac{3}{\tilde{\sigma}_p+2} r\cos\theta,~\Phi^0_m = -\left(r + \frac{\tilde{\sigma}_p-1}{\tilde{\sigma}_p+2}\frac{1}{r^2}\right)\cos\theta,
\end{equation}
where $\tilde{\sigma}_p$ is the dimensionless particle conductivity $\tilde{\sigma} = \sigma_p/\sigma_0$. The $O(\epsilon)$ electrohydrodynamic problem in the spherical coordinates is
\begin{eqnarray}
	&&\left( \frac{\partial^2}{\partial r^2} + \frac{2}{r}\frac{\partial}{\partial r} + \frac{\cot\theta}{r^2}\frac{\partial}{\partial\theta} + \frac{1}{r^2}\frac{\partial^2}{\partial\theta^2} - \frac{1}{r^2 \sin^2\theta}\right) \omega = -H, \label{unbounded1} \\
	&&\omega = \frac{1}{r}\left[\frac{\partial}{\partial}(r u_\theta) - \frac{\partial u_r}{\partial \theta}\right], \label{unbounded2} \\
	&& \frac{1}{r^2}\frac{\partial}{\partial r}(r^2 u_r) + \frac{1}{r\sin\theta}\frac{\partial}{\partial\theta}(\sin\theta u_\theta) = 0 \label{unbounded3},
\end{eqnarray}
where $u_r$ and $u_\theta$ are the velocity components and $\omega$ is the vorticity. \refeq{unbounded1} is derived by taking the curl of the Stokes equation and $H$ is the curl of the Coulomb force $\bff$,
\begin{equation}
	H = \frac{1}{r}\left[\frac{\partial}{\partial r}(rf_\theta) - \frac{\partial f_r}{\partial \theta}\right]. \nonumber
\end{equation}
Due to the symmetry, we can simplify the domain to be the quarter of the plane, $r\in(r,+\infty)$ and $\theta \in (0,\pi/2)$. Boundary conditions are
\begin{eqnarray}
	&& u_\theta = \omega = 0,~\frac{\partial u_r}{\partial \theta} = 0 ~{\rm at}~ \theta = 0,\pi/2, \nonumber\\
	&& u_r = u_\theta = 0,~\omega = \frac{\partial v_\theta}{\partial r} ~{\rm at}~ r = 1,\nonumber\\
	&& u_r \to 0,~u_\theta \to 0,~\omega \to 0 ~{\rm as}~ r\to\infty. \nonumber
\end{eqnarray}
The problem, \refeq{unbounded1} - \refeq{unbounded3} is solved numerically with the Chebyshev collocation method \citep{trefethen2000spectral}. In practical numerical implementation, \refeq{unbounded1} and \refeq{unbounded2} are combined to eliminate the vorticity $\omega$ and corresponding boundary conditions.

\bibliographystyle{jfm}

\end{document}